\documentstyle[prd,preprint,aps]{revtex}
\newcommand{\newc}{\newcommand}
\newc{\gsim}{\lower.7ex\hbox{$\;\stackrel{\textstyle>}{\sim}\;$}}
\newc{\lsim}{\lower.7ex\hbox{$\;\stackrel{\textstyle<}{\sim}\;$}}
\newc{\gev}{\,{\rm GeV}}
\newc{\mev}{\,{\rm MeV}}
\newc{\ev}{\,{\rm eV}}
\newc{\kev}{\,{\rm keV}}
\newc{\tev}{\,{\rm TeV}}
\newc{\mz}{m_Z}
\newc{\mpl}{M_{Pl}}
\newc{\chifc}{\chi_{{}_{\!F\!C}}}
\newc\order{{\cal O}}
\newc\CO{\order}
\newc\CL{{\cal L}}
\newc\CY{{\cal Y}}
\newc\CH{{\cal H}}
\newc\CM{{\cal M}}
\newc\CF{{\cal F}}
\newc\CD{{\cal D}}
\newc\CN{{\cal N}}
\newc{\eps}{\epsilon}
\newc{\re}{\mbox{Re}\,}
\newc{\im}{\mbox{Im}\,}
\newc{\invpb}{\,\mbox{pb}^{-1}}
\newc{\invfb}{\,\mbox{fb}^{-1}}
\newc{\yddiag}{{\bf D}}
\newc{\yddiagd}{{\bf D^\dagger}}
\newc{\yudiag}{{\bf U}}
\newc{\yudiagd}{{\bf U^\dagger}}
\newc{\yd}{{\bf Y_D}}
\newc{\ydd}{{\bf Y_D^\dagger}}
\newc{\yu}{{\bf Y_U}}
\newc{\yud}{{\bf Y_U^\dagger}}
\newc{\ckm}{{\bf V}}
\newc{\ckmd}{{\bf V^\dagger}}
\newc{\ckmz}{{\bf V^0}}
\newc{\ckmzd}{{\bf V^{0\dagger}}}
\newc{\X}{{\bf X}}
\newc{\bbbar}{B^0-\bar B^0}

\newc{\sgn}{\mbox{sgn}\,}
\newc{\m}{{\bf m}}
\newc{\msusy}{M_{\rm SUSY}}
\newc{\munif}{M_{\rm unif}}
\newc{\slepton}{{\tilde\ell}}
\newc{\Slepton}{{\tilde L}}
\newc{\sneutrino}{{\tilde\nu}}
\newc{\selectron}{{\tilde e}}
\newc{\stau}{{\tilde\tau}}
\relax
\def\beq{\begin{equation}}
\def\eeq{\end{equation}}
\def\bea{\begin{eqnarray}}
\def\eea{\end{eqnarray}}
\newc{\ie}{{\it i.e.}}          \newc{\etal}{{\it et al.}}
\newc{\eg}{{\it e.g.}}          \newc{\etc}{{\it etc.}}
\newc{\cf}{{\it c.f.}}
\def\dis{\displaystyle}
\def\beq{\begin{equation}}
\def\eeq{\end{equation}}
\def\[{\left[}
\def\]{\right]}

\def\dis{\displaystyle}
\def\beq{\begin{equation}}
\def\eeq{\end{equation}}
\def\[{\left[}
\def\]{\right]}

\input epsf

\def\Dsl{\,\raise.15ex\hbox{/}\mkern-13.5mu D} %this one can be subscripted
\def\delsl{\raise.15ex\hbox{/}\kern-.57em\partial}
\def\Ksl{\hbox{/\kern-.6000em\rm K}}
\def\Asl{\hbox{/\kern-.6500em \rm A}}
\def\Qsl{\hbox{/\kern-.6000em\rm Q}}
\def\gradsl{\hbox{/\kern-.6500em$\nabla$}}
\def\bar#1{\overline{#1}}

\begin{document}
%\begin{titlepage}
\draft
%\begin{flushright}
%IFUAP-HEP-03-01\\
%July 2003\\
%\end{flushright}
\title{The decays $h^{\pm} \to W^{\pm} h^{0}(\mbox{a}^0)$ within an
extension of the MSSM with one complex Higgs triplet}
\author{E. Barradas-Guevara$^{(a)}$, O. F\'elix-Beltr\'an$^{(b)}$,
J. Hern\' andez-S\' anchez$^{(c)}$ and  A. Rosado$^{(d)}$}
\address{$^{(a)}$ Fac. de Cs. F\'{\i}sico-Matem\'aticas, BUAP.
Apdo. Postal 1364, C.P. 72000 Puebla, Pue., M\'exico\\
$^{(b)}$ Instituto de F\'{\i}sica, UNAM, Apdo. Postal 20-364,
M\'exico 01000 D.F., M\'exico\\
$^{(c)}$ Centro de Investigaci\'on Avanzada en Ingenier\'{\i}a 
Industrial, Universidad Aut\'onoma del Estado de Hidalgo,
Carretera Pachuca-Tulancingo Km. 4.5, Ciudad Universitaria, C.P. 42020
Pachuca, Hgo., M\'exico\\
$^{(d)}$ Instituto de F\'{\i}sica, BUAP.
Apdo. Postal J-48, C.P. 72570 Puebla, Pue., M\'exico}

\date{\today}

\maketitle

\begin{abstract}
\noindent The vertices $h^{\pm} W^{\mp} h^{0}$ and $h^{\pm} W^{\mp}
\mbox{a}^{0}$, involving the gauge bosons $W^\mp$, the lightest charged
($h^{\pm}$), the lightest CP-even neutral ($h^{0}$), and the lightest CP-odd
neutral ($\mbox{a}^{0}$) Higgs bosons, arise  within the context of many
extensions of the SM, and they can be used to probe the Higgs sector of
such extensions via the decays $h^{\pm} \to W^{\pm} \, h^{0}(\mbox{a}^{0})$.
We discuss the strength of these vertices for an extension of the MSSM with
an additional complex Higgs triplet. By using this model, we find regions
of the parameter space where the decays $h^{\pm} \to  W^{\pm} \, h^{0}
(\mbox{a}^{0})$ are not only kinematically allowed, but they also become
important decay modes and in some cases the dominant ones, with
$BR(h^{\pm} \to  W^{\pm} \, h^{0}) \approx BR(h^{\pm} \to  W^{\pm} \,
\mbox{a}^{0})$.
\end{abstract}

%\end{titlepage}
\setcounter{footnote}{0}
\setcounter{page}{2}
\setcounter{section}{0}
\setcounter{subsection}{0}
\setcounter{subsubsection}{0}

%%%%%%%%%%%%%%%%%%%%%%%%%%%%%%%%%%%%%%%%%%%%%%%%%%%%%%%%%%%%%%%%%%%%%%%

\narrowtext

\subsection{Introduction}
%%%%%%%%%%%%%%%%%%%%%%%%%%%%%%%%%%%%%%%%%%%%%%%%%%%%%%%%%%%%%%%%%%%%

The Higgs spectrum of many well motivated extensions of the Standard Model
(SM) often include charged Higgs bosons whose detection in future colliders
would constitute a clear evidence of a Higgs sector beyond the minimal SM
\cite{stanmod,kanehunt}. In particular, the Two-Higgs-Doublet-Model (THDM)
has been extensively studied as a prototype of a Higgs sector that includes
two charged Higgs bosons ($H^\pm$)\cite{kanehunt}, however, a definitive
test of the mechanism of electroweak symmetry breaking will require further
studies of the complete Higgs spectrum. In addition, probing the properties
of charged Higgs bosons through their decays could help find out whether they
are indeed associated with a weakly-interacting theory, as in the case of the
popular minimal supersymmetric extension of the SM (MSSM)\cite{susyhix}, or
with a strongly-interacting scenario \cite{stronghix}. Furthermore, these
tests should also allow to probe the symmetries of the Higgs potential, and
to determine whether the charged Higgs bosons belong to a weak-doublet or to
some larger multiplet.

Decays of charged Higgs bosons have been studied in the literature, 
including the radiative modes $W^{\pm}\gamma, W^{\pm}Z^0$ \cite{hcdecay}, 
mostly within the context of the THDM or its MSSM incarnation and, more
recently, for the effective Lagrangian extension of the THDM
\cite{ourpaper}. Charged Higgs bosons production at hadron colliders was
studied long ago \cite{ldcysampay} and, recently, more systematic
calculations of production processes at the future Large Hadron Collider
(LHC) have been presented \cite{newhcprod}. Current bounds on the mass of
the charged Higgs bosons can be obtained at Tevatron, by studying the top
decay $t \to bH^+$, which already eliminates some regions of the parameter
space \cite{lhcbounds}, whereas LEP-2 bounds give approximately
$m_{H^{+}} > 80$ $GeV$ \cite{lepbounds}.

On the other hand, the vertex $H^{\pm} W^{\mp} h^0$ deserves special
attention because it can give valuable information about the underlying
structure of the gauge and scalar sectors. In the first place, the decay
mode $H^{\pm} \to W^{\pm} h^0$ might be detected at the LHC as
claimed in Ref.\cite{hcwhdetect}, within the context of the MSSM.
Furthermore, the vertex $H^{\pm} W^{\mp} h^0$ can also induce the associated
production of $H^{\pm} \, h^0$ at hadron colliders, through a virtual
$W^{\pm \, *}$ in the s-channel which would become a relevant production
mechanism for heavy charged Higgs bosons. In this paper we are interested
in studying the strength of this important vertex for an extension of the
MSSM with one additional complex Higgs triplet (OHT-MSSM) \cite{trip,esquia}, via
the decay $h^{\pm}_k \to W^{\pm} \, h^{0}$, with $h^{\pm}$ the lightest
charged and $h^{0}$ the lightest neutral Higgs bosons of the model.

This article is organized as follows: in section B, we present and discuss
briefly the  results for the branching ratio (BR) of the charged Higgs boson
decay in the context of the MSSM, we include in our numerical calculations
the leading order radiative corrections. In section C, we discuss the
strength of the vertex for an extended supersymmetric model that includes a
complex Higgs triplet (OHT-MSSM). We perform a numerical analysis to search
for values of the Higgs boson masses that allow for the decay $H^{\pm}
\to W^{\pm} h^0$. Finally, we summarize  our conclusions in section D.

%%%%%%%%%%%%%%%%%%%%%%%%%%%%%%%%%%%%%%%%%%%%%%%%%%%%%%%%%%%%%%%%%%%%%%%
\subsection{The vertex $H^{\pm} W^{\mp} h^0$ in the MSSM}
%%%%%%%%%%%%%%%%%%%%%%%%%%%%%%%%%%%%%%%%%%%%%%%%%%%%%%%%%%%%%%%%%%%%%%%

The simplest model that predicts charged Higgs bosons is the MSSM, which
includes two scalar doublets of equal hypercharge, namely,
$\Phi_1=(\phi^+_1,\phi^0_1)$ and $\Phi_2=(\phi^+_2,\phi^0_2)$. Besides two
charged Higgs bosons ($H^\pm$), the spectrum of the MSSM includes two neutral
CP-even states ($h^0,H^0$, with $m_{h^0} < m_{H^0}$), as well as a neutral
CP-odd state ($A^0$).
Diagonalization of the charged mass matrices gives the expression for
the charged Higgs boson mass-eigenstates: $H^{\pm}=\cos\beta \, \phi^{\pm}_1
+ \sin\beta \, \phi^{\pm}_2$, where $\tan\beta(=v_2/v_1)$ denotes the ratio
of v.e.v.'s of each doublet.

\bigskip

\noindent {\it B.1 The decay $H^{\pm} \to W^{\pm} \, h^0$ in the  MSSM
with radiative corrections}.

Whenever kinematically allowed, the vertex $H^{\pm} W^{\pm} h^0$ could induce
the decay $H^{\pm} \to W^{\pm} \, h^0$. For the light SM-like Higgs boson,
this decay is proportional to the factor $\cos^2 (\beta -\alpha )$, which
determines its strength.

Other relevant decays of the charged Higgs boson are the modes into fermion
pairs, which include the decays $H^{+(-)} \to \bar{\tau}\nu_\tau, \, c
\bar{b} \, (\tau \bar{\nu}_\tau, \, \bar{c} b)$, and possibly into $t \bar{b}
\, (\bar{t} b)$. If the charged Higgs bosons are indeed associated with the
Higgs mechanism, their couplings to fermions should come from the Yukawa
sector and the corresponding decays should have a larger BR for the modes
involving the heavier fermions. The latter could be tested in a simple way if
a comparison of the modes $H^{+(-)} \to \bar{\tau}\nu_\tau \, (\tau \bar{\nu}
_{\tau})$ and $H^{+(-)} \to \bar{\mu} \nu_\mu \, (\mu \bar{\nu}_{\mu})$ led
to very different BR's.

The masses of the two CP-even neutral Higgs bosons ($h^0,H^0$) and the
charged pair ($H^\pm$) are conveniently determined in terms of the mass
of the CP-odd state ($A^0$) and $\tan\beta$. In the MSSM the quartic
couplings are given in terms of the gauge couplings, which implies that the
light neutral Higgs boson must satisfy the (tree-level) bound $m_{h^0} \leq
\cos 2 \beta \, m_Z$. However, this relation is modified by important
corrections arising from top/stop loops, which result into a bound $m_{h^0}
\lesssim 130$ $GeV$ \cite{partdat}.

In the decoupling limit ($m_A \gg m_Z$) the parameters of the potential lead
to the relation: $\cos^2(\beta -\alpha ) \simeq m_Z^2/m_{A^0}^2 $, which
remains small for large values of $m_{A^0}$. One also obtains an approximately
degenerate spectrum of heavy Higgs bosons, {\it i.e.} $m_{H^{\pm}} \simeq
m_{H^0} \simeq m_{A^0}$, while the mixing angles satisfy the following
relation: $\alpha \simeq \beta-\pi/2$. Therefore, in the context of the MSSM,
only the decay mode $W^{\pm} h^0$ is allowed for most regions of the
parameter space. We have performed a detailed parametric search for contour
regions for the branching ratio of $H^\pm \to W^\pm \, h^0$, by using the
program HDECAY \cite{hdecay}, our results are shown in Fig. 1.
%By
%taking $\tan \beta \approx 10 \,(30)$ one obtains a typical BR of the order
%of $10^{-2}$ ($10^{-5}$) for $m_{H^{\pm}} \approx 300$ $GeV$.

%%%%%%%%%%%%%%%%%%%%%%%%%%%%%%%%%%%%%%%%%%%%%%%%%%%%%%%%%%%%%%%%%%%%%%%
\subsection{The vertex $h^{\pm} W^{\mp} h^0$ in a SUSY model with an
additional complex Higgs triplet}
%%%%%%%%%%%%%%%%%%%%%%%%%%%%%%%%%%%%%%%%%%%%%%%%%%%%%%%%%%%%%%%%%%%%%%%

The supersymmetric model with two doublets and a complex triplet (OHT-MSSM)
\cite{trip,esquia} is one of the simplest extensions of the minimal supersymmetric
standard model (MSSM) that allows to study phenomenological consequences of
an explicit breaking of the custodial symmetry SU(2) \cite{esquia}. 

\bigskip

\noindent {\it C.1 The Higgs sector of the model}. The model includes
two Higgs doublets and a (complex) Higgs triplet given by
\begin{equation}
\Phi_1 = \left( 
\begin{array}{c}
{\phi_1}^0 \\ 
{\phi_1}^-
\end{array}
\right) \,\,\, , \,\,\, \Phi_2= \left( 
\begin{array}{c}
{\phi_2}^+ \\ 
{\phi_2}^0
\end{array}
\right) \,\,\, , \,\,\,
\sum = \left( 
\begin{array}{cc}
\sqrt{\frac{1}{2}} \xi^0 & - \xi_2^+ \\ 
\xi_1^- & - \sqrt{\frac{1}{2}} \xi^0
\end{array}
\right) \,\,\, .
\end{equation}
The Higgs triplet is described in terms of a $2 \times 2$ matrix
representation; $\xi^0$ is the complex neutral field, and $\xi_1^-, \,
\xi_2^+$ denote the charged scalars.  The most general gauge invariant
and renormalizable superpotential that can be written for the Higgs
superfields $\Phi _{1,2}$ and $\Sigma $ is given by:
\begin{equation}
W=\lambda \Phi _{1}\cdot \Sigma \Phi _{2}+\mu _{D}\Phi _{1}\cdot \Phi
_{2}+\mu _{T}\mbox{Tr}(\Sigma ^{2})\,\,\,,
\end{equation}
where we have used the notation $\Phi _{1}\cdot \Phi _{2}\equiv \epsilon
_{ab}\Phi _{1}^{a}\Phi _{2}^{b}$.
The resulting scalar potential involving only the Higgs fields 
is thus written as 
$$
V=V_{SB}+V_{F}+V_{D}\,\,\,, 
$$
\noindent where $V_{SB}$ denotes the most general soft-supersymmetry breaking
potential \cite{trip}. In turn, the full scalar potential can be split into
its neutral and charged parts, {\it i.e.}, $V = V_{charged} + V_{neutral}$.

Besides the supersymmetry-breaking mass terms, $m_i^2$ ($i = 1,\,2,\,3$),
the potential depends on the parameters $\lambda, \,\, \mu_D, \,\, \mu_T,
\,\, A, \,\, B$. For simplicity, we will assume that there is no CP violation
in the Higgs sector, and thus, all the parameters and the v.e.v.'s are
assumed to be real. The explicit expression of the Higgs potential is given
in Ref.\cite{trip}.

We can also combine the v.e.v.'s of the Higgs doublet as $v_{D}^{2} \equiv
v_{1}^{2}+v_{2}^{2}$ and define $\mbox{tan}\beta \equiv {v_{2}}/{v_{1}}$.
Furthermore, the parameters $v_{D}$, $v_{T}$, $m_W^2$ and $m_Z^2$ are
related as follows:
\begin{equation}
\begin{array}{cl}
m_{W}^{2}= & \frac{1}{2}g^{2}(v_{D}^{2}+4v_{T}^{2}), \\ 
m_{Z}^{2}= & {\displaystyle{\frac{{\frac{1}{2}g^{2}v_{D}^{2}}}{{\mbox{cos}
^{2}{\theta }_{W}}}}}\,\, ,
\end{array}
\end{equation}
which implies that the $\rho$-parameter is different from 1 at the tree
level, namely,
\begin{equation}
\rho \equiv \frac{M_{W}^{2}}{M_{Z}^{2}\mbox{cos}^{2}\theta _{W}}
=1+4R^{2},\,\,\,\,\,\,R\equiv \frac{v_{T}}{v_{D}}\,\,\,.
\end{equation}
The bound on $R$ is obtained from the $\rho $ parameter, which lies in the
range 0.9993 - 1.0006, from the global fit reported in Refs.\cite{partdat,trizzo}.
Thus, $R \leq 0.012$ and $v_{T}\leq 3 \, \, GeV$. We
have taken into account this bound in our numerical analysis.

\bigskip

\noindent {\it C.2 Mass spectrum}. Diagonalization of the mass matrices (and
the resulting mass eigenvalues) and mixing matrix will allow us to analyze
the coupling $ H^{\pm}_k W^{\mp} H^0_j$ ($k=1,\,2,\,3$, and $j=1,\,2,\,3$)
and the coupling $ H^{\pm}_k W^{\mp} A^0_j$ ($k=1,\,2,\,3$, and $j=1,\,2$).
The CP-even (odd) mass eigenstates are denoted by $H^0_1$, $H^0_2$ and
$H^0_3$ ($A^0_1$ and $A^0_2$), ordered according to their masses,
$m_{H^0_1} < m_{H^0_2} < m_{H^0_3}$ ($ m_{A^0_1} < m_{A^0_2}$).
The charged Higgs states are denoted by $H^{\pm}_{k}$ with $ m_{H^{\pm}_{1}}
< m_{H^{\pm}_{2}} < m_{H^{\pm}_{3}}$. We will denote the lightest charged
scalar $H^{\pm}_1$, the lightest neutral scalar $H^0_1$ and the lightest
neutral pseudoscalar $A^0_1$ as $h^{\pm}$, $h^0$ and $\mbox{a}^0$,
respectively. Because of the large number of
parameters appearing in our model, which include $\tan \beta$, $R$,
$\lambda$, $\mu _{D}$, $\mu _{T}$, $A$, $B_{D}$, and $B_{T}$, it is
convenient to consider only a few simple cases. In each, we will try to
identify useful relations or trends for the behavior of the Higgs boson
masses and couplings. In order to perform the numerical analysis leading
to the allowed regions in the parameter space and the Higgs boson
masses, we will make the following asumptions: a) $\tan \beta $ is an
independent variable; b) $R$ takes the representative value 0.01; c)
$\lambda$ takes the value 0.5; and d) the remaining parameters will cover
the regions allowed by SUSY. Specifically, we will consider charged Higgs
bosons masses in the range 100 - 300 $GeV$. Furthermore, we will restrict
our numerical analysis to the following specific scenarios (which were
introduced and discussed in Ref.\cite{trip}):

\noindent - Scenario I: $B_D=\mu _D=0$, which represents the scenario when
the spontaneous symmetry breaking (SSB) is dominated by the effects of the
Higgs triplet, where we will consider the following cases: A) $B_T=\mu _T=
A$; B) $B_T=\mu _T=-A$; C) $B_T=-\mu _T=A$; D) $-B_T=\mu _T=A$.

\noindent - Scenario II: $B_T=\mu_T=0$. In this scenario the SSB is dominated
by the effects of the Higgs doublets, where the following cases will be
considered: A) $B_D=\mu _D=A$; B) $B_D=\mu_D=-A$; C) $B_D=-\mu_D=A$;
D) $-B_D=\mu_D=A$.

\noindent - Scenario III: $\left| B_D \right| =\left| B_T \right| =
\left| \mu _D \right| = \left| \mu _T \right| =\left| A \right|$. Both
doublets and the triplet contribute to the SSB. Within this scenario
several cases are considered: for instance A) $B_D= B_T=\mu _D=\mu_T=A$, as
well as 15 other combinations with positive and negative signs.

For each point in the parameter space, within the above scenarios, we
will determine the allowed regions by requiring the scalar squared mass 
eigenvalues to be positive and the Higgs potential lying in a global 
minimum. In these allowed regions, the masses of the physical Higgs bosons
contained in the model are computed numerically.  

\bigskip

\noindent {\it C.3 The vertex $H^{\pm}_{k} W^{\mp} h^0$,
$H^{\pm}_{k} W^{\mp} \mbox{A}^0$ and $H^{\pm}_{k} W^{\mp} Z^0$}
($k=1,\,2,\,3$). We consider only the cases of the lightest
neutral CP-even scalar, $h^0$ and the lightest neutral CP-odd scalar,
$\mbox{a}^0$. We will use the expression for the vertex $H^{\pm}_{k}W^{\mp}h^0$
and $H^{\pm}_{k}W^{\mp}\mbox{a}^0$ for the OHT-MSSM reported in Refs.
\cite{trip,esquia}. To present a complete study of the branching ratios of
the charged Higgs bosons, we also discuss the vertex
$H^{\pm}_{k} W^{\mp} Z^0$, which could dominate in some specific scenarios.

By using the expression for the rotation matrices of the charged and
neutral Higgs sectors, $U$ and $V$, we can write the coefficient of the
vertex $H^{\pm}_k W^{\mp}h^0$ and $H^{\pm}_k W^{\mp}\mbox{a}^0$, namely
$\eta_k^{h^0}$ and $\eta_k^{\mbox{a}^0}$, respectively, as follows:
\begin{equation}
\eta_k^{h^{0}}= i \left( \dis{\frac{1}{\sqrt{2}}}(V^S_{11} U_{2(k+1)} -
V^S_{21}  U_{1(k+1)})+
\dis{\frac{1}{4}}V^S_{31}( U_{4(k+1)} - U_{3(k+1)}) \right) \, ,
\end{equation}
\noindent and 
\begin{equation}
\eta_k^{\mbox{a}^0}= -i \left( \dis{\frac{1}{\sqrt{2}}}(V^{PS}_{11}
U_{2(k+1)} - V^S_{21}  U_{1(k+1)})+
\dis{\frac{1}{4}}V^{PS}_{31}( U_{4(k+1)} - U_{3(k+1)}) \right) \, ,
\end{equation}
\noindent where $H^{\pm}_{k}$ denote the charged Higgs bosons of the model,
and $h^0$($\mbox{a}^0$) corresponds to the lightest neutral scalar(pseudoescalar)
Higgs boson of the model. The $U_{jk}$'s denote the elements of the matrix,
which relates the physical charged Higgs bosons ($H^+_1,H^+_2,H^+_3$) and the Goldstone
boson $G^+_0$ (which gives mass to the $W^+$) with the fields:
$\phi_2^+$, $\phi_1^-{}^{*}$, $\xi_2^+$ and $\xi_1^-{}^{*}$, as follows:
\begin{eqnarray}
\left ( 
\begin{array}{r}
\phi_2^+ \\
\phi_1^-{}^{*} \\
\xi_2^+ \\
\xi_1^-{}^{*} \\
\end{array} \right ) \ =  \ 
\left ( 
\begin{array}{rrrr}
U_{11} & U_{12} & U_{13}  & U_{14} \\
U_{21} & U_{22} & U_{23}  & U_{24} \\
U_{31} & U_{32} & U_{33}  & U_{34} \\
U_{41} & U_{42} & U_{43}  & U_{44} \end{array} \right ) 
\left ( 
\begin{array}{r}
G^+ \\
H_1^+ \\
H_2^+ \\
H_3^+ \\
\end{array} \right ) \ .
\end{eqnarray}
The $V^S_{ij}$'s and the $V^{PS}_{ij}$'s denote the elements of the
rotation matrix for the CP-even and CP-odd neutral sector, respectively.
The matrices $V^S$ and $V^{PS}$ relate the physical scalars
$(H_1^0,H_2^0,H_3^0)$, the physical pseudoscalars ($A^0_1,A^0_2$) and
Goldstone boson $G^0$ (which gives mass to the $Z^0$), with the real and
imaginary parts of the fields $\phi_1^0$, $\phi_2^0$, $\xi^0$, in the
following way.
\begin{eqnarray}
\left ( 
\begin{array}{r}
\sqrt{\frac{1}{2}} \, \mbox{Re}(\phi_1^0) \\
\sqrt{\frac{1}{2}} \, \mbox{Re}(\phi_2^0) \\
\sqrt{\frac{1}{2}} \, \mbox{Re}(\xi^0) \\
\end{array} \right ) \ =  \ 
\left ( 
\begin{array}{rrr}
V^{S}_{11} & V^{S}_{12} & V^{S}_{13} \\
V^{S}_{21} & V^{S}_{22} & V^{S}_{23} \\
V^{S}_{31} & V^{S}_{32} & V^{S}_{33} \end{array} \right ) 
\left ( 
\begin{array}{r}
H_1^0 \\
H_2^0 \\
H_3^0 \\
\end{array} \right ) \ .
\end{eqnarray}

\noindent and

\begin{eqnarray}
\left ( 
\begin{array}{r}
\sqrt{\frac{1}{2}} \, \mbox{Im}(\phi_1^0) \\
\sqrt{\frac{1}{2}} \, \mbox{Im}(\phi_2^0) \\
\sqrt{\frac{1}{2}} \, \mbox{Im}(\xi^0) \\
\end{array} \right ) \ =  \ 
\left ( 
\begin{array}{rrr}
V^{S}_{11} & V^{PS}_{12} & V^{PS}_{13} \\
V^{S}_{21} & V^{PS}_{22} & V^{PS}_{23} \\
V^{S}_{31} & V^{PS}_{32} & V^{PS}_{33} \end{array} \right ) 
\left ( 
\begin{array}{r}
A_1^0 \\
G^0 \\
A_2^0 \\
\end{array} \right ) \ .
\end{eqnarray}

On the other hand, in this model the vertex $H^{\pm}_k W^{\mp} Z^0$ is also
induced at tree level due to violation of the custodial symmetry.
The expression for the vertex $H^{\pm}_k W^{\mp} Z^0$ is given by
\begin{equation}
H_{k}^{\pm}W^{\mp}_{\mu}Z^0_{\nu}: {\pm}i \, g^2 v_T (U_{3(k+1)} - U_{4(k+1)})
\, \cos\theta_W \, g_{\mu\nu} .
\end{equation}
\noindent  One can see than only the triplet components contribute to 
this vertex, while the dependence on $v_T$ gives a suppression effect. 

\bigskip

\noindent {\it C.4 Branching ratios for the principal two and three body
decay modes of $h^{\pm}$}.  We now discuss
the BR for the charged Higgs bosons, including the decay widths of the
dominant modes of $h^{\pm}$, which turn out to be the following: 
1) $h^{\pm} \to W^{\pm} Z^0$; 2) $h^{\pm} \to W^{\pm}h^0$;
3) $h^{\pm} \to W^{\pm}\mbox{a}^0$; 4) $h^{+(-)} \to t \bar{b} \, (\bar{t} b)$;
5) $h^{+(-)} \to \bar{\tau} \nu_{\tau} \, (\tau \bar{\nu})$.
In order to discuss the BR for the charged Higgs bosons in the low
mass region $100 \, GeV < m_{h^{\pm}} < 200 \, GeV$, it is necessary to
include the dominant modes of the three-body decay of $h^{\pm}$, namely:
6) $h^{\pm} \to Z^0 W^{\pm *} \to Z^0 f \bar{f'}$;
7) $h^{\pm}  \to h^0 W^{\pm *} \to h^0 f \bar{f'}$;
8) $h^{\pm}  \to \mbox{a}^0 W^{\pm *} \to \mbox{a}^0 f \bar{f'}$ (It has been shown that
this decay is a potentially strong tree-level process in the THDM-I
\cite{akeroyd1,akeroyd2});
9) $h^{+} \to t^* \bar{b} \to W^+ b \bar{b}$.
The decay widths for each of the above modes are given as \cite{zerwas}:
\begin{enumerate}
\item The decay $h^{\pm} \to W^{\pm} Z^0$:
\begin{eqnarray}
\Gamma \left( h^{\pm} \to W^{\pm} Z^0 \right) &=& 
g^2 \, v^2_T |(U_{32} - U_{42})|^2 \, \cos^2\theta_W {\lambda}^{1/2}
(1, \kappa_W, \kappa_Z)
\nonumber\\
& &\left( \frac{(m^2_{h^{\pm}}-m^2_W-m^2_Z)^2 +8m^2_W m^2_Z}
{64 \pi m^2_Z m^2_W m_{h^{\pm}}} \right)
\end{eqnarray}
\noindent Here, $\kappa_W=m^2_W/m^2_{h^{\pm}}$ and
$\kappa_Z=m^2_Z/m^2_{h^{\pm}}$, and $\lambda^{1/2}$ is the usual
kinematic factor
\begin{equation}
\lambda^{1/2}(a,b,c)= \sqrt{(a-b-c)^2 - 4bc}.
\end{equation}

\item The decay $h^{\pm}  \to W^{\pm} h^0$:
\begin{equation}
\begin{array}{cl}
\Gamma \left( h^{\pm} \to W^{\pm}h^0 \right) = &
\dis{\frac{g^2 {\lambda}^{1/2}(m_{h^{\pm}}^{2},m_{W}^{2},m_{h^0}^{2})}
{64 \pi m_{h^{\pm}}^3}} {\mid \eta_1^{h^{0}} \mid}^2  \\
 & \times \[ m_{W}^{2}- 2 (m_{h^{\pm}}^{2} + m_{h^0}^{2}) + 
 \dis{\frac{{(m_{h^{\pm}}^{2} - m_{h^0}^{2})}^{2}}{m_{W}^{2}}} \],
\end{array}
\end{equation}
\noindent This decay is proportional to the factor $|\eta_1^{h^{0}}|^2$.

\item The decay $h^{\pm}  \to W^{\pm} \mbox{a}^0$:
\begin{equation}
\begin{array}{cl}
\Gamma \left( h^{\pm} \to W^{\pm}\mbox{a}^0 \right) = &
\dis{\frac{g^2 {\lambda}^{1/2}(m_{h^{\pm}}^{2},m_{W}^{2},m_{\mbox{a}^0}^{2})}
{64 \pi m_{h^{\pm}}^3}} {\mid \eta_1^{\mbox{a}^0} \mid}^2  \\
 & \times \[ m_{W}^{2}- 2 (m_{h^{\pm}}^{2} + m_{\mbox{a}^0}^{2}) + 
 \dis{\frac{{(m_{h^{\pm}}^{2} - m_{\mbox{a}^0}^{2})}^{2}}{m_{W}^{2}}} \],
\end{array}
\end{equation}
\noindent This decay is proportional to
the factor $|\eta_1^{\mbox{a}^0}|^2$. In the MSSM the two-body decay of
the charged Higgs boson into $W^{\pm} A^0$ is kinematically not allowed.

\item The decay $h^{+(-)} \to t \bar{b} \, (\bar{t} b)$:
\begin{equation}
\begin{array}{c}
\Gamma (h^{+(-)} \to t \bar{b} \, (\bar{t} b)) = 
\dis{\frac{3g^2}{32 \pi m_{W}^{2} m_{h^{\pm}}^3}} 
\lambda^{1/2} ( m_{h^{\pm}}^2, m_{t}^{2},  m_{b}^{2}) \\  
\times \left[ (m_{h^{\pm}}^2 - m_{t}^{2}- m_{b}^{2})
(m_{b}^{2} {\tan}^2 \beta + m_{t}^{2} {\cot}^2 \beta )
- 4m_{b}^{2}m_{t}^{2} \right].
\end{array} 
\end{equation}

\item The decay $h^{+(-)} \to \bar{\tau} \nu_{\tau} \,
(\tau \bar{\nu}_{\tau})$:
%\begin{equation}
%\begin{array}{cl}
%\Gamma (h^{+(-)} \to \bar{\tau} \nu_{\tau} \, (\tau \bar{\nu}_{\tau})) =
%&\dis{\frac{g^2}{32 \pi m_{W}^{2} m_{h^{\pm}}^2}} \lambda^{1/2}\left( 
%m^2_{h^{\pm}}, 0,  m_{\tau}^{2} \right) \\  
%& \times \, m_{\tau}^{2} \, {\tan}^2 \beta \, (m_{h^{\pm}}^2
%- m_{\tau}^{2}).
%\end{array}
%\end{equation}
\begin{equation}
\Gamma (h^{+(-)} \to \bar{\tau} \nu_{\tau} \, (\tau \bar{\nu}_{\tau})) =
\dis{\frac{g^2 \, m_{\tau}^{2} \, {\tan}^2 \beta \,}{32 \pi m_{W}^{2}
m_{h^{\pm}}^3}} \, (m_{h^{\pm}}^2 - m_{\tau}^{2}) \,
\lambda^{1/2}( m^2_{h^{\pm}}, 0,  m_{\tau}^{2}).\\
\end{equation}

\item The decay $h^{\pm} \to Z^0 W^{\pm *} \to Z^0 f \bar{f'}$:
\begin{equation}
\Gamma \left( h^{\pm} \to Z^0 W^{\pm *} \to Z^0 f \bar{f'} \right) =
\mid F_Z \mid ^2 \, \frac{3 \, g^4 \, m_{h^{\pm}}}{512 \, \pi^3} \,
F(m_Z/m_{h^{\pm}})
\end{equation}
\noindent with $F_Z = (g \, v_T/ m_W) (U_{32} - U_{42}) \, \cos\theta_W$ and 
\begin{eqnarray}
F(x) &=& - \mid 1 - x^2 \mid \left( \frac{47}{2} x^2 - \frac{13}{2} +
\frac{1}{x^2} \right) -3 ( 1 - 6 x^2 + 4 x^4) \mid \ln x \mid \nonumber\\
& &+ 3 \frac{1 - 8 x^2 + 20 x^4}{\sqrt{4 x^2 -1}}
\cos^{-1} \left( \frac{3 x^2 - 1}{2 x^3} \right). \nonumber
\end{eqnarray}
\noindent We have simplified the expression for this width by taking
the following approximation for the $W$ propagator:
$[(P-k)^2 - m_W^2]^{-1} \approx [m^2_{h^{\pm}}- 2P \cdot k]^{-1}$,
where $P^{\mu}$ and $k^{\mu}$ are the four momenta of the $H^{\pm}$
and $Z^0$ bosons, respectively.

\item The decay $h^{\pm}  \to h^0 W^{\pm *} \to h^0 f \bar{f'}$:
\begin{equation}
\Gamma \left( h^{\pm}  \to h^0 W^{\pm *} \to h^0 f \bar{f} \right) = 
\frac{9 \, g^4}{256 \, \pi^3} \, \mid \eta_1^{h^0} \mid ^2 
\, m_{h^{\pm}} \, G_{h^0 W^{\pm}} \\
\end{equation}
\noindent where
\begin{eqnarray}
G_{ij} &=& \frac{1}{4} \left\{ 2(-1+\kappa_j-\kappa_i) \sqrt{\lambda_{ij}}
\left[ \frac{\pi}{2} + \arctan \left( \frac{\kappa_j(1-\kappa_j+\kappa_i)
-\lambda_{ij}}{(1-\kappa_i)\sqrt{\lambda_{ij}}}
\right) \right] \right. \nonumber\\
&&\left. +(\lambda_{ij}-2\kappa_i)\log(\kappa_i)+\frac{1}{3}(1-\kappa_i)
\left[5(1+\kappa_i)-4\kappa_j-\frac{2}{\kappa_j} \lambda_{ij} \right]
\right\} 
\end{eqnarray}
\noindent and
\begin{equation}
\lambda_{ij}=-1+2\kappa_i+2\kappa_j-(\kappa_i-\kappa_j)^2,
\end{equation}
\noindent with $\kappa_i=m^2_i/m^2_{h^{\pm}}.$

\item The decay $h^{\pm}  \to \mbox{a}^0 W^{\pm *} \to \mbox{a}^0 f \bar{f'}$:
\begin{equation}
\Gamma \left( h^{\pm}  \to h^0 W^{\pm *} \to h^0 f \bar{f} \right) = 
\frac{9 \, g^4}{256 \pi^3} {\mid \eta_1^{\mbox{a}^0} \mid}^2 m_{h^{\pm}}
G_{\mbox{a}^0 W^{\pm}} \\
\end{equation}
\noindent The coefficient $G_{\mbox{a}^0 W^{\pm}}$ has been defined in eqs.(17,18)

\item The decay $h^{+} \to t^* \bar{b} \to W^+ b \bar{b}$:
\begin{eqnarray}
\Gamma (h^{+} \to t^* \bar{b} \to W^+ b \bar{b}) &=& 
\frac{1}{2} \, K_{H^{\pm}tb}
\left\{ \frac{\kappa_W^2}{\kappa_t^3}(4\kappa_W\kappa_t) + 3\kappa_t
-4\kappa_W) \log \left( \frac{\kappa_W(\kappa_t-1)}{\kappa_t-\kappa_W} \right)
\right. \nonumber\\
&&+ (3\kappa_t^2 - 4\kappa_t - 3\kappa_W^2 + 1) \log
\left( \frac{\kappa_t-1}{\kappa_t-\kappa_W} \right) - \frac{5}{2}
\nonumber\\
&&\left. + \frac{1-\kappa_W}{\kappa_t^2} (3\kappa_t^2 - \kappa_t\kappa_W
- 2\kappa_t\kappa_W^2 + 4\kappa_W^2)) + \kappa_W \left( 4 -
\frac{3}{2}\kappa_W \right)
\right\}
\end{eqnarray}
\noindent with
\begin{equation}
K_{h^{\pm} tb}=\frac{3\,g^4\,m_t^4}{1024\,\pi^3\,m^4_W} \frac{1}
{\tan^2\beta} m_{h^{\pm}}.
\end{equation}
\end{enumerate}

Finally, we evaluate numerically the BR for these nine principal modes,
using the expressions given in Eqs. (9-21). In our computations, we will
take $\sin^2\theta_W=0.223$, $m_W=80.4$ $GeV$, $m_b=4.5$ $GeV$, $m_t=174.3$
$GeV$ \cite{partdat}. We present in Figs. 2-10 our
results in some specific scenarios, which can be summarized as follows:

\noindent - In scenario I the numerical calculations are performed for
$\lambda = 0.5$ and several values of $\tan\beta$. We consider case A
within this scenario for the lightest charged Higgs boson $h^{\pm}$.
We show our results for $\tan\beta= 15$, 30 and 50, in  Figs. 2, 3 and 4,
respectively.
In this scenario, $h^{\pm} \to W^{\pm}h^0$ and $h^{\pm} \to
W^{\pm}\mbox{a}^0$ are the dominat decays for $\tan\beta= 15$. For
$\tan\beta= 30$, the decays $h^{\pm} \to W^{\pm}h^0$ and $h^{\pm} \to
W^{\pm}\mbox{a}^0$ are important modes, but $h^{\pm} \to  W^{\pm}Z^0$
becomes the dominant decay. For $\tan\beta= 50$, $h^{\pm} \to W^{\pm}h^0$
and $h^{\pm} \to W^{\pm}\mbox{a}^0$ are still important decays (BR of the
order of $10^{-1}$), but we can see that $W^{\pm}Z^0$ is the dominant mode
for $m_{h^{\pm}} < 195 \, GeV$ and $t \bar{b} \, (\bar{t} b)$ is the
dominant mode for $m_{h^{\pm}} > 200 \, GeV$.

\noindent - In scenario II, where it is mimicked the MSSM, we consider the
case D. Taking $\lambda = 0.5$, we plot our results for $\tan\beta= 5$, 15
and 30, in  Figs. 5, 6 and 7, respectively. In this scenario,
$h^{\pm} \to W^{\pm}h^0$ and $h^{\pm} \to W^{\pm}\mbox{a}^0$ are important
decays for $140 \, GeV < m_{h^{\pm}} < 160 \, GeV$, becoming the dominat
decays for $\tan\beta \approx 5$.

\noindent - Finally, for scenario III we consider the case F ($B_D= B_T=-A;
\; \mu _D=\mu_T=A$). Here, both doublets and tripet contribute equally to
the SSB. We calculate the BR's for the principal modes by taking
$\lambda = 0.5$ and some values of $\tan\beta$. Our results are shown in
Fig. 8 (for $\tan\beta = 5$), Fig. 9 (for $\tan\beta = 15$) and Fig. 10
(for $\tan\beta = 15$). The behavior of the BR of the different decay modes
is similar to that observed in scenario II.A.

To end this section, we want to point out the following. It is
clear that for the OHT-MSSM there are regions in the parameter space that
correspond to either the dominant ($BR \approx 1$) or moderate ($10^{-2}
\lesssim BR \lesssim 10^{-1}$) case. Therefore, the observation of the decays
$h^{\pm}\to W^{\pm} \, h^{0}(\mbox{a}^{0})$, as the dominant modes, would
back up the OHT-MSSM. On the other hand, the moderate case could arise either
of the MSSM or the OHT-MSSM. The observation of charged Higgs bosons in the
region of the parameter space predicted by the MSSM would not discard the
OHT-MSSM, while the detection of several charged Higgs bosons would correspond
to a model with a more elaborate Higgs sector (such as Higgs triplets).

%%%%%%%%%%%%%%%%%%%%%%%%%%%%%%%%%%%%%%%%%%%%%%%%%%%%%%%%%%%%%%%%%%%%%%%
\subsection{Conclusions}
%%%%%%%%%%%%%%%%%%%%%%%%%%%%%%%%%%%%%%%%%%%%%%%%%%%%%%%%%%%%%%%%%%%%%%%

We have studied the charged Higgs vertices $h^{\pm} W^{\mp} h^{0}(\mbox{a}^0)$,
within the context of an extension of the minimal supersymmetric standard
model (MSSM) with an additional complex Higgs triplet (OHT-MSSM) and then we
have analyzed the decays $h^{\pm} \to W^{\pm} \, h^{0}(\mbox{a}^0)$ in the
frame of this model. We found regions in the parameter space where the decays
$h^{\pm} \to W^{\pm} \, h^{0}(\mbox{a}^0)$, are not only kinematically
allowed, but they also become important decay modes and in some cases the
dominant decay modes, with $BR(h^{\pm} \to W^{\mp} \, \mbox{a}^0) \approx
BR(h^{\pm} \to W^{\mp} \, h^{0})$. We conclude that for the OHT-MSSM there
are regions in the parameter space that correspond to the case when the
$W^{\pm}h^0(\mbox{a}^0)$ decay modes are dominant or gets a BR in the range
$10^{-2} - 10^{-1}$ (moderate case). The detection of the decay $h^{\pm}\to
W^{\pm}h^{0}(\mbox{a}^0)$, as the dominant modes, would favor the SUSY
triplet case. On the other hand, the moderate case could arise either of the
MSSM or the OHT-MSSM. The detection of charged Higgs bosons in the region of
the parameter space predicted by the MSSM would not discard the OHT-MSSM.
Clearly, the observation of several charged Higgs bosons would correspond to
a model with a more elaborate Higgs sector, such as the OHT-MSSM.

\bigskip

%%%%%%%%%%%%%%%%%%%%%%%%%%%%%%%%%%%%%%%%%%%%%%%%%%%%%%%%%%%%%%%%%%%%%%%%%%
Acknowledgments: This work was supported in part by CONACYT (Mexico).
During this work, O.F-B. was supported by the UNAM (Mexico) under Proyecto
DGAPA-PAPIIT IN116202-3. The work of J.H.S. was suported by Programa de
Consolidaci\'on Institucional-Conacyt (M\'exico) and by a SEP-PROMEP grant.
A.R. acknowledges financial support by SNI (Mexico). We have enjoyed
fruitful and enlighten discussions with Lorenzo D\'{\i}az-Cruz.
%%%%%%%%%%%%%%%%%%%%%%%%%%%%%%%%%%%%%%%%%%%%%%%%%%%%%%%%%%%%%%%%%%%%%%%%%%%

\newpage

\begin{center}
FIGURE CAPTIONS
\end{center}
\noindent{\bf Fig. 1.}
BR $(H^{\pm} \to W^{\pm} h^0)$ in the MSSM, with radiative corrections
to the Higgs mass as included in HDECAY, with $m_{\widetilde{q}}=500$ $GeV$,
$\mu = 100$ and $A_0=1500$.

\bigskip

\noindent{\bf Fig. 2.} Branching ratios of the charged Higgs bosons
$h^{\pm}$ decaying into the principal modes for scenario I (case A),
considering $\lambda = 0.5$. The various line drawings correspond to
the different modes:
(1) $h^{\pm} \to W^{\pm} Z^0$;
(2) $h^{\pm} \to W^{\pm}h^0$;
(3) $h^{\pm} \to W^{\pm}\mbox{a}^0$;
(4) $h^{+(-)} \to t \bar{b} \, (\bar{t} b)$;
(5) $h^{+(-)} \to \bar{\tau} \nu_{\tau} \,
(\tau \bar{\nu})$;
(6) $h^{\pm} \to Z^0 W^{\pm *} \to Z^0 f \bar{f'}$;
(7) $h^{\pm}  \to h^0 W^{\pm *} \to h^0 f \bar{f'}$;
(8) $h^{\pm}  \to \mbox{a}^0 W^{\pm *} \to \mbox{a}^0 f \bar{f'}$;
(9) $h^{+} \to t^* \bar{b} \to W^+ b \bar{b}$.
These modes are shown for the lightest charged Higgs boson,
for $\tan\beta =15$.

\bigskip

\noindent{\bf Fig. 3.} Same as in Fig. 2, but for $\tan\beta = 30$.

\bigskip

\noindent{\bf Fig. 4.} Same as in Fig. 2, but for $\tan\beta = 50$.

\bigskip

\noindent{\bf Fig. 5.} Same as in Fig. 2, but for Scenario II (case D),
for $\tan\beta = 5$.

\bigskip

\noindent{\bf Fig. 6.} Same as in Fig. 5, but for $\tan\beta = 15$.

\bigskip

\noindent{\bf Fig. 7.} Same as in Fig. 5, but for $\tan\beta = 30$.

\bigskip

\noindent{\bf Fig. 8.} Same as in Fig. 2, but for Scenario III (case F),
for $\tan\beta = 5$.

\bigskip

\noindent{\bf Fig. 9.} Same as in Fig. 8, but for $\tan\beta = 15$.

\bigskip

\noindent{\bf Fig. 10.} Same as in Fig. 8, but for $\tan\beta = 30$.


\bigskip

\newpage

\begin{references}
%1
\bibitem{stanmod} S. L. Glashow, Nucl. Phys. {\bf 22}, 579 (1961); S.
Weinberg, Phys. Rev. Lett. {\bf 19}, 1264 (1967); A. Salam, Proc. 8th NOBEL
Symposium, ed. N. Svartholm (Almqvist and Wiksell, Stockholm, 1968), p. 367.

%2
\bibitem{kanehunt}
S. Dawson {\it et al.}, The Higgs Hunter's Guide,
2nd ed., Frontiers in Physics Vol. {\bf 80}
(Addison-Wesley, Reading MA, 1990).

%3
\bibitem{susyhix}
M. Carena {\it et al.}, Report of the Tevatron Higgs
working group; FERMILAB-CONF-00-279-T; hep-ph/0010338; see also:
C. Balazs {\it et al.}, Phys. Rev. {\bf D59}, 055016 (1999);
J.L. D\'{\i}az-Cruz {\it et al.}, Phys. Rev, Lett. {\bf 80}, 4641 (1998).

%4
\bibitem{stronghix} See for instance:
B. Dobrescu, Phys. Rev. {\bf D63}, 015004 (2001).

%5
\bibitem{hcdecay}
J. L. D\'{\i}az-Cruz and M.A. P\' erez, Phys. Rev. {\bf D33}, 273 (1986);
J. Gunion, G. Kane and J. Wudka, Nucl. Phys. {\bf B299}, 231 (1988);
A. Mendez and A. Pomarol, Nucl. Phys. {\bf B349}, 369 (1991);
E. Barradas et al., Phys. Rev. {\bf D53}, 1678 (1996);
M. Capdequi Peyranere, H. E. Haber, and P. Irulegui,
Phys. Rev. {\bf D44}, 191 (1991);
S. Kanemura, Phys. Rev. {\bf D61}, 095001 (2000);
{\it The Higgs Working Group: Summary Report} to appear in the Proceedings
of the Workshop on Physics at TeV Colliders. Les Houches, France 2003,
hep-ph/0406152;
J. Hern\'andez-S\'anchez et al., Phys. Rev. {\bf D69}, 095008 (2004).

%6
\bibitem{ourpaper}
J. L. D\'{\i}az-Cruz, J. Hern\'andez-S\'anchez and
 J.J. Toscano, Phys. Lett. {\bf B512}, 339 (2001).

%7
\bibitem{ldcysampay}
J. L. D\'{\i}az-Cruz and O.A. Sampayo, Phys. Rev. {\bf D50}, 6820 (1994);
J. Gunion et al., Nucl. Phys. {\bf B294}, 621 (1987);
M. A. P\' erez and A. Rosado, Phys. Rev. {\bf D30}, 228 (1984).

%8
\bibitem{newhcprod} M. Bisset {\it et al.}, Eur. Phys. J. {\bf C19}, 143 (2001);
A. Barrientos Bendezu and  B.A. Kniehl, Phys. Rev. {\bf D63}, 015009 (2001);
Z. Fei et al., Phys. Rev. {\bf D63}, 015002 (2001);
D. J. Miller  et al., Phys. Rev. {\bf D61}, 055011 (2000).

%9
\bibitem{lhcbounds} F. Abe {\it et al.} (CDF Collaboration), Phys. Rev. Lett. 
{\bf 79}, 357 (1997).

%10
\bibitem{lepbounds} For a review see:
F. Borzumati and A. Djouadi, Phys. Lett. {\bf B549}, 170 (2002);
D. P. Roy, Mod. Phys. Lett. {\bf A19}, 1813 (2004).

%11
\bibitem{hcwhdetect} S. Moretti, Phys. Lett. {\bf B481}, 49 (2000);
K. A. Assamagan, Y. Coadou and A. Deandrea,
Eur. Phys. J. {\bf C4}, 9 (2002).

%12
\bibitem{trip}
Olga F\'elix-Beltr\'an, Int. Jour. of Mod. Phys. {\bf A17}, 465 (2002).

%13
\bibitem{esquia}  
J. R. Espinosa and M. Quir\'{o}s, Nucl. Phys. {\bf B384}, 113 (1992).

%%
%\bibitem{jfg} 
%J. F. Gunion and H. E. Haber,
%Phys. Rev. {\bf D67}, 075019 (2003). 
%
%%
%\bibitem{npb380} 
%J. L. Diaz Cruz and A. Mendez,
%Nucl. Phys.{\bf B380}, 39 (1992). 

%14
\bibitem{partdat} K. Hagiwara {\it et al.}, ({\it Particle Data Group}),
Phys. Rev. {\bf D66}, 010001 (2002).

%
%\bibitem{cotaneu}  
%Review of Particle Physics,
%Phys. Rev. {\bf D66}, 309 (2002).

%15
\bibitem{hdecay}
A. Djouadi, J. Kalinowski and M. Spira, Comput. Phys. Commun.
{\bf 108}, 56 (1998).

%16
\bibitem{trizzo}  
T. Rizzo, Mod. Phys. Lett. {\bf A6}, 1961 (1992).

%17
\bibitem{akeroyd1} A.G. Akeroyd, Nucl. Phys. {\bf B544}, 557 (1999).

%18
\bibitem{akeroyd2} A.G. Akeroyd, A. Arhrib and E. Naimi, Eur. Phys. J.
{\bf C12}, 451 (2000).

%19
\bibitem{zerwas} Analytical expressions for the two and three body decay
modes of SUSY Higgs bosons can be found in A. Djouadi, J. Kalinowski and
P. Zerwas, Z. Phys. {\bf C70}, 435 (1996).


\end{references}
\end{document}